\documentclass{iau} 
\usepackage{graphicx}
\usepackage{wrapfig}

\title[Open our eyes to wider fields in VLBI surveys] 
{Open our eyes to wider fields in VLBI surveys}

\author[Krist{\'o}f Rozgonyi \& S{\'a}ndor Frey]   
{Krist{\'o}f Rozgonyi$^1$ \and S{\'a}ndor Frey$^{2,3}$}

\affiliation{$ ^1 $Department of Physics of Complex Systems, E\"otv\"os Lor\'and University, P\'azm\'any P\'eter s\'et\'any 1/A, H-1117 Budapest, Hungary \\ e-mail: \texttt{rstofi@gmail.com} \\[\affilskip] $ ^2 $F\"OMI Satellite Geodetic Observatory, PO Box 585, H-1592 Budapest, Hungary \\[\affilskip] $ ^3 $Konkoly Observatory, MTA CSFK, PO Box 67, H-1525 Budapest, Hungary \\ e-mail: \texttt{sandor.frey@gmail.com}}

\pubyear{2016}
\volume{324}  
\jname{New Frontiers in Black Hole Astrophysics}
\editors{A.C. Editor, B.D. Editor \& C.E. Editor, eds.}
\begin{document}

\maketitle

\begin{abstract}

The observation and imaging of hundreds or thousands of radio sources with the technique of very long baseline interferometry (VLBI) is a computationally intensive task. However, these surveys allow us to conduct statistical investigations of large source samples, and also to discover new phenomena or types of objects. The field of view of these high-resolution VLBI imaging observations is typically a few arcseconds at cm wavelengths. For practical reasons, often a much smaller fraction of the field, the central region is imaged only. With an automated process we imaged the $\sim$1.5-arcsec radius fields around more than 1000 radio sources, and found a variety of extended radio structures. Some of them are yet unknown in the literature.

\end{abstract}

\firstsection 
\section{Treasure-house of data: VIPS}

We examined the VLBA Imaging and Polarimetry Survey (VIPS, \cite[Helmboldt et al. 2007]{Helmboldt2007}). It contains 1127 AGN observations taken at 5~GHz with the Very Long Baseline Array (VLBA). The calibrated VLBI visibility data sets can be downloaded from the VIPS data collection index page\footnote{\texttt{http://www.phys.unm.edu/$\sim$gbtaylor/VIPS/vipscat/vipsncapindx.shtml}}. The original field of view of the interferometer was $\sim$1.5 arcsec but the imaging was typically preformed only in the central area of 128~mas\,$\times$\,128~mas. Even though large structures were reported in the original publication and subsequent studies (\cite[Helmboldt et al. 2007]{Helmboldt2007}, \cite[Tremblay et al. 2016]{Tremblay2016}), we could find further objects in the full-field images with structures extending beyond the central area.

\section{Imaging of the extended field}

The large sample size and the extended fields motivated us to image the sources and analyse the data in an automated way. The imaging and model fitting were performed through standard procedures using Difmap (\cite[Shepherd et al. 1994]{Shepherd94}). We modeled the visibility data with a central elliptical Gaussian component. In addition, extended structures were fitted with circular Gaussians. The number of these components was determined by the signal-to-noise ratio (SNR) in the residual map. We applied a 6-$\sigma $ threshold for fitting a new component, as done in \cite[Helmboldt et al. (2007)]{Helmboldt2007}. Thereafter, we calculated the separation of the circular components from the central elliptical Gaussian. We marked the sources with at least one fitted component farther away than 50~mas from the centre as candidates for having large-scale structure within the full VLBA field of view.

\section{Results and plans for the future}

\begin{figure}[h]
\begin{center}
 \includegraphics[width=3.5in]{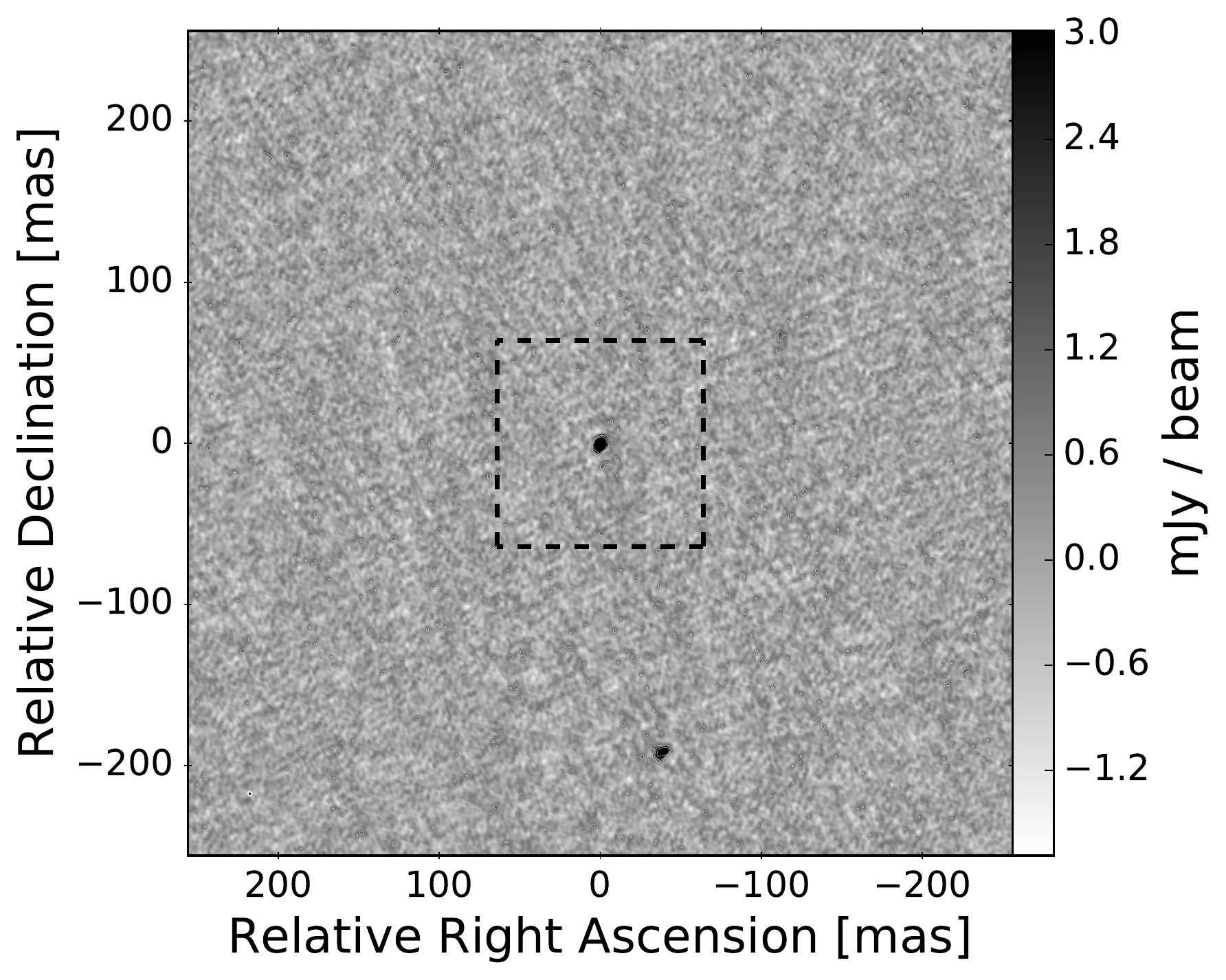} 
 \caption{5-GHz VLBI image of J15048+5438. A compact component is seen $\sim$200 milli-arcseconds (mas) south of the central component. Follow-up multi-frequency observations could be useful to decide if this is a kpc-separation dual radio AGN, a gravitationally lensed background source with two images, or an AGN core plus a hot spot embedded in an extended lobe that is resolved out with the VLBA. The area originally imaged in VIPS is enclosed by the dashed line.}
   \label{fig1}
\end{center}
\end{figure}

We found 60 sources as candidates with extended structure among the 1127 VIPS sources. To check the quality of the candidate sample, we visually inspected all the wide-field images. We found that in a few cases the distant components are unreliable because of the low flux density of the fitted components. An increased SNR threshold in the automated model-fitting procedure would eliminate these questionable cases. Further quantitative classification of the sample is needed (and planned). Published information from the literature, analysis of archival data, and eventually new targeted observations would be necessary to reveal the true nature of these objects, on a case-by-case basis. These sources show great diversity in morphology and brightness. The large, typically $\sim$0.1--1~kpc structures could most probably be extended jet structures. However, the possibility of dual radio AGN or gravitationally lensed radio sources should also be investigated. As an illustration, we present an example from the candidate sample (Fig.\,\ref{fig1}). 

We plan to perform the same automated procedure with other publicly available survey data, in a hope to find further objects with VLBI structure extended beyond what is already known. 

\begin{acknowledgments}
This work was supported by the Hungarian National Research, Development and Innovation Office (OTKA NN110333). KR thanks the \'UNKP-16-2 New National Excellence Program of the Ministry of Human Capacities for support.
\end{acknowledgments}

\end{document}